\documentclass[letterpaper,  journal, twoside,twocolumn]{support/IEEEtran}
\usepackage{amsmath}
\usepackage{times}
\usepackage[pdftex]{graphicx}
\usepackage{subfigure}
\usepackage{amsmath,amssymb,amsopn,amstext,amsfonts}
\usepackage{cancel}
\usepackage[noadjust]{cite}
\usepackage{soul}
\usepackage{caption}
\usepackage{gensymb}
\usepackage[ruled,linesnumbered]{algorithm2e}
\usepackage{booktabs} 

\usepackage{balance}
\usepackage{color}
\usepackage{mathtools}
\usepackage{bm}
\usepackage{diagbox}
\usepackage{float}
\usepackage{epstopdf}
\usepackage{url}
\usepackage{multirow}
\usepackage{tikz}
\usepackage{subeqnarray}
\usepackage{cases}
\usepackage{booktabs}
\usepackage[linkcolor=black,citecolor=black,urlcolor=black,colorlinks=true]{hyperref}
\usepackage{algpseudocode}
\newtheorem{myTheo}{Theorem}
\newtheorem{lemma}{Lemma} 

\newtheorem{remark}{Remark}

\newtheorem{assumption}{Assumption}

\soulregister\cite7
\soulregister\citep7
\soulregister\citet7
\soulregister\ref7
\soulregister\it7
\soulregister\pageref7

\graphicspath{{figures/}}
\DeclareGraphicsExtensions{.pdf,.png,.jpg,.eps}
\IEEEoverridecommandlockouts

\title{\LARGE \bf Data-Driven Leader-following Consensus for Nonlinear Multi-Agent Systems against Composite  Attacks: A Twins Layer Approach}

\author{
  \vskip 1em
  {Xin Gong, \emph{Member, IEEE},
  Jintao Peng,
  Dong Yang,
  Zhan Shu, \emph{Senior Member, IEEE},
  Tingwen Huang, \emph{Fellow, IEEE},~and~Yukang Cui, \emph{Member, IEEE}
  }

  \thanks{
    This work was partially supported by the National Natural Science Foundation of China under Grant 61903258, Guangdong Basic and Applied Basic Research Foundation 2022A1515010234 and the Project of Department of Education of Guangdong Province 2022KTSCX105. 

X. Gong is with the Department of Mechanical Engineering, The University of Hong Kong, Pokfulam Road, Hong Kong (e-mail: {\tt\small gongxin@connect.hku.hk}).

J. Peng and Y. Cui  are with the College of Mechatronics and Control Engineering, Shenzhen University, Shenzhen 518060, China (e-mail: {\tt\small cuiyukang@gmail.com; pengjintao98@163.com}).

D. Yang is with China Academy of Space Technology, Beijing 100094, China (e-mail: {\tt\small qbdyzy@sina.com}).

Z. Shu is with the Department of Electrical and Computer
Engineering, University of Alberta, Edmonton, AB T6G 2R3, Canada (e-mail: {\tt\small hustd8@gmail.com}).

T. Huang is with Texas A\&M University at Qatar, Doha 23874, Qatar (e-mail: {\tt\small tingwen.huang@qatar.tamu.edu}).

  }
}

\begin{document}
  \maketitle
  \begin{abstract}
This paper studies the leader-following consensuses of uncertain and nonlinear multi-agent systems against composite attacks (CAs), including Denial of Service (DoS) attacks and actuation attacks (AAs). 
A double-layer control framework is formulated, where a digital twin layer (TL) is added beside the traditional cyber-physical layer (CPL), inspired by the recent Digital Twin technology. Consequently, the resilient control task against CAs can be divided into two parts: 
One is distributed estimation against DoS attacks on the TL and the other is resilient decentralized tracking control against actuation attacks on the CPL. 
First, a distributed observer based on switching estimation law against DoS is designed on TL.
Second,
a distributed model free adaptive control (DMFAC) protocol based on attack compensation against AAs is designed on CPL.

Moreover, the uniformly ultimately bounded convergence of consensus error of the proposed double-layer DMFAC algorithm is strictly proved. Finally, the simulation verifies the effectiveness of the resilient double-layer control scheme.
\end{abstract}
 
\section{Introduction}
\IEEEPARstart{I}{n} recent years, multi-agent systems (MASs) have gained popularity and are used in many different contexts, including satellite formation\cite{satellite_formation_01,satellite_formation_02}, mobile robots \cite{mobile_robots_01,mobile_robots_02}, autonomous surface vehicles \cite{ASV_01,ASV_CC_01} and UAV cluster control \cite{UAV_01,UAV_CC_01} , etc. 
Cooperative control of MASs has become a hot research fields\cite{ASV_CC_01,MAS_XG_BC_01,MAS_XG_01,UAV_CC_01} and the consensus problem is one of the most important research focuses in Cooperative control. 
In this paper, we work toward a resilient control schemes of leader-following consensus control of nonlinear MASs against various cyber attacks in the framework of hierarchical distributed control.

Multi-agent systems rely on mutual communication to achieve cooperative control. 
However, in a complex communication environment, it cannot avoid communication uncertainty, such as cyber attacks. 
Common types of cyber attacks include Denial of Service (DoS) attacks \cite{MAS_DoS_01,MAS_DoS_FTC_01} and actuation attacks (AAs) \cite{CPS_AAS_01,MAS_attack_analysis_01,AFTC_AAS_01,MAS_AAs_RC_01}. 
Among them, DoS attacks could cut off the communication channel between agents through various means while AAs directly inject attack signals into the actuators of agents to offset the system control input. 
All of the above brings damage to the security, robustness and information integrity of MASs, making it difficult to design controllers to resist the damage and achieve great self-control. 
In the existing papers, defense strategies against cyber attacks are mainly based on attack detection and attack adaptive methods. 
The former needs to constantly detect and identify attacks, which brings a great computational burden to the system. 
The latter can achieve acceptable system performance without attack detection, but its defense strategy is only effective for one type of attack according to existing papers \cite{AFTC_AAS_01,MAS_MA_RC_01,MAS_AAs_RC_01}. 
In this paper, the idea of hierarchical control \cite{HC_01} is introduced to cope with multiple cyber attacks at the same time through building a digital twin layer. 
And we focus on leader-following consensus control for nonlinear discrete systems with unknown models. 
In practical applications, model uncertainty and nonlinear problems can not be avoided, hence control of nonlinear unknown systems cannot be ignored.

Most of the research on the consensus control of MASs assumes that the system dynamics are known and have accurate dynamic models. 
However, an accurate system model means high measurement cost, so that system model in practical application is imprecise and uncertain from the perspective of cost saving. 
Meanwhile, nonlinearity is a typical nature of complexity in nature and even the linear dynamics of agents cannot avoid the existence of nonlinear parts. 
Therefore, the research on consensus control of nonlinear uncertain MASs is of great significance. 
In dealing with system nonlinear problems, neural networks (NNs) is a feasible approach known from the papers \cite{NMAS_NNS_LFC_01,NmobileA_NNs_secondC_01,NMAS_NNs_AC_TD_01} because of its excellent estimation ability, but the adaptive controller design based on neural network needs a training process to provide appropriate training parameters. 
In addition, a consensus control approach based on iterative learning control (ILC) is also being studied for nonlinear multi-agent systems \cite{MAS_ILC_C_01,NMAS_ILC_FC_01,NMAS_ILC_C_01}. However, this method is based on the assumption that there is a priori knowledge of the leader's state trajectory in the whole control process. 
Therefore, it is not real-time leader-following tracking. 
Fortunately, the model free adaptive control (MFAC) method is used to achieve real-time leader-following consensus control for discrete-time nonlinear systems with unknown dynamics. 
In the MFAC \cite{NCAN_ILC_MFAC_27,NMAS_ILC_MFAC_29,NMAS_MAFC_DoS_C_34!!}, nonlinear systems were represented by a linear data model linked to input/output (I/O) data, and the control protocol was created based on the linear data model without the need for an accurate understanding of the system structure.

There are also many papers applying NNs and ILC to MFAC \cite{NMAS_ILC_MFAC_01,NMAS_ILC_MFAC_CT_42,NMAS_ILC_MFAC_29,NolnearS_NNS_MFAC_FC_38,NNCS_MAFC_NNs_FC_DoS_41}.
In \cite{NolnearS_NNS_MFAC_FC_38,NNCS_MAFC_NNs_FC_DoS_41}, the MFAC is used to solve the problem of imprecise model and NNs is used to estimate the sensor error. 
The network computation in these papers is set up in the cloud due to high computing burden, but there is also a hidden danger of cyber attacks. 
And in paper \cite{NMAS_ILC_MFAC_29}, ILC is applied to iteratively learn the optimal control input sequence in the whole control process according to data-driven models, but it is also unavoidable to assume that the expected trajectory or leader state is known throughout the control period, so that real-time tracking cannot be realized.
In contrast to apply NNs and ILC to MFAC above, this paper introduces the idea of hierarchical control in MFAC, in which we build a digital twin layer (TL), corresponding to the cyber-physical layer (CPL). 
TL has the same number of leaders and followers and the same topology as CPL. 
Since the twin layer has no practical physical significance, it has high confidentiality and security and can be immune to AAs. 
Therefore, we can divide the leader-following control against DoS attacks and AAs into two parts: 
resisting DoS attacks to achieve consensus control on the TL and resisting AAs to achieve consensus control on the CPL. 
A switching control law is designed for DoS on the TL and an adaptive controller with attacks compensation is designed on the CPL for unbounded AAs. 
Both of them achieve the uniformly ultimately bounded (UUB) convergence of tracking error.

Inspired by the foregoing discussions, a new hierarchical control scheme based on MFAC is proposed to realize the leader-following consensus control for nonlinear MASs against DoS attacks and AAs. The main contributions of the article are as follows:
\begin{enumerate}
\item A double-layer resilient control framework, including TL and CPL, is designed to achieve resilient leader-following consensus against cyber attacks. The adding TL can be deployed in the Cloud, which is not existing physically and has less physical meaning. Thus, the TL has higher security and confidentiality than the CPL, which is immune from AAs. Consequently, the control task can be divided into two parts: distributed estimation against DoS attacks on the TL and decentralized control against AAs on the CPL. The attack defense strategy based on the DMFAC approach is designed for TL and CPL, respectively.

\item On the TL, a distributed switching estimation scheme is proposed, which switches according to whether the DoS attacks occur or not. The above estimation scheme owns UUB convergence, with an explicit upper bound. The tolerable DoS attack magnitude is also discussed.

\item The considered AAs on the CPL can be unbounded, which outperforms most of the previous works towards bounded AAs \cite{}. Based on the rationale of compact form dynamic linearization (CFDL), an MFAC-based decentralized control scheme against unbounded AAs is designed, which possesses UUB convergence.
\end{enumerate}


\emph{Notations:}
The symbols $\mathbb{R}^n$ and $\mathbb{R}^{n\times n}$ refer to sets of real vectors of dimension $n$ and matrix of dimension $n \times n$ respectively. And symbol $\mathbb{N}$ refer to nonnegative integer set. Denote $I_n\in \mathbb{R}^{n \times n}$ as an identity matrix of dimension $n \times n$ and $1_n\in \mathbb{R}^n$ as a column vector filled with 1. $\otimes$ represents the Kronecker product. $\rm{diag}(x_1,x_2,\ldots,x_i)$ refer to a diagonal matrix with $x_1,x_2,\ldots,x_i$ as the diagonal elements.
\section{ Preliminaries And System Setup}\label{section2}
\subsection{Graph Theory}\label{graph}
In Graph Theory, a directed graph $\mathcal{G}=(\mathcal{V},\mathcal{E}, \mathbb{A})$ can be used to illustrate information communication between agents, where $\mathcal{V}=\{1,2,\ldots,N\}$ is a group of agents and $\mathcal{E}=\mathcal{V}\times \mathcal{V}$ is a group of edges indicating the flow of information between agents. An edge $e_{ij}$ in $\mathcal{G}$ indicates that the information of node $j$ is available to that of node $i$, and agent $j$ is denoted as a neighbor of agent $i$. The index set of all neighbors of agent $i$ is denoted by $N_i=\{ j:(i,j)\in \mathcal{E} \}$. In an undirected graph, $(i,j)\in\mathcal{E}\Leftrightarrow(j,i)\in\mathcal{E}$. The adjacent matrix $\mathbb{A}\triangleq [a_{i j}]\in \mathbb{R}^{N\times N}$, where $a_{i j}=1$ if $(i,j)\in \mathcal{E}$, and $a_{i j}=0$ otherwise.  The Laplacian matrix $L\triangleq [l_{ij}]\in \mathbb{R}^{N\times N}$, where $l_{ii}=\sum_{j=1,j\neq  i}^N a_{ij}$, and $l_{ij}=-a_{ij}$ for $i\neq j$. It is easy to obtain that the accumulation of elements in every row of matrix is zero and has $N$ nonzero eigenvalue. A information channel between agent $i$ and agent $j$ is a sequence of edges $(i,j_1),(j_2,j_3),\ldots,(j_l,j)$ in $\mathcal{G}$ with distinct agents $j_k$, $k=1,2,\ldots,l$. If there exists a path for every two nodes, we said the undirected graph is strongly connected. 

In this paper, $\mathcal{G}=(\mathcal{V},\mathcal{E}, \mathbb{A})$ and Laplacian matrix $L$ describe the topological relationship between followers. And here is a matrix $C={\rm diag}(c_1,c_2,\cdots,c_N)$ to indicate whether leaders communicate with followers, where $c_i=1$ if agent $i$ can receive information from leader, and $c_i=0$ otherwise.
\subsection{CFDL Data Models}
 In the framework of formation-tracking control, we consider an unknown nonlinear MAS, in which the agents be classified into two groups:
\begin{itemize}
  \item [1)]  One leader is denoted as agent $0$;
 \item [2)] $N$ followers are denoted as agent $1,2,\ldots,N$.  
\end{itemize}

 In existing research, a group of agents with the same dynamic are frequently taken into account in consensus control.
As opposed to that, the MASs considered in this paper is heterogeneous and even unknown nonlinear, and the nonlinear dynamics of leader $0$ and $N$ followers are shown below:
  \begin{equation} \label{eq_sys}
  \begin{cases}
      y_0(k+1)=f_0(y_0(k))\\
      y_i(k+1)=f_i(y_i(k),u_i(k)),\quad i=1,2,\ldots,N,
  \end{cases}
  \end{equation}
  where $y_i(k)\in \mathbb{R}$ is the state output, $u_i(k)\in \mathbb{R}$ is control input and $f_i(\cdot)$ is an unidentified nonlinear function of agent $i$, respectively.

As described in \ref{graph}, $\mathcal{G}=(\mathcal{V},\mathcal{E}, \mathbb{A})$ only represents the topological relationship between followers. Considering the communication between leaders and followers, $\mathcal{\bar{G}}=(\mathcal{V}\cup \{0\},\mathcal{\bar{E}}, \mathbb{\bar{A}})$ is introduced to represent the topological relationship between all agents in MASs. The following is the necessary assumption for $\mathcal{\bar{G}}$.

  \begin{assumption}\label{as_1}
  The communication graph $\mathcal{\bar{G}}$ which describes information communication among agents is directed and fixed strongly connected, that is, at least one of the follower agents has access to the leader.
  \end{assumption}
  
  \begin{remark}
  $\mathcal{\bar{G}}$ is fixed strongly connected in Assumption \ref{as_1}, which ensures that information can be transmitted between any two followers. 
  As long as one follower receives the leader's state information, all followers can get the leader's state information through the topological network. $\hfill \hfill \square $
  \end{remark} 

Because the system is nonlinear even unknown, we can not get a definite linear system model. 
So a new model free adaptive control method based on data-driven is introduced, in which only the control input and control output of the system are used in the control process. 
The MFAC has been widely used in nonlinear or unknown systems, and the following necessary assumptions for the MFAC are put forward for completing the following  analysis.

\begin{assumption}(\rm\cite{NMAS_MAFC_C_31})\label{as_2}
The partial derivative of $f_i(\cdot)$ with respect to $u_i(k)$ is continuous.
\end{assumption}

\begin{assumption}(\rm\cite{NMAS_MAFC_C_31})\label{as_3}
The follower systems in \eqref{eq_sys} satisfies the generalized Lipschitz condition, that is, if $|\Delta u_i(k)|\neq 0$, $ | {\Delta y_i(k+1) |} \leq b_c  |\Delta u_i(k) |$ holds for any $k$, where $\Delta y_i(k+1)=y_i(k+1)-y_i(k)$, $\Delta u_i(k)=u_i(k)-u_i(k-1)$ and $b_c$ is positive constant.
\end{assumption}

\begin{remark}
Assumption \ref{as_2} is a conventional constraint condition for nonlinear systems. 
Assumption \ref{as_3} states that the bounded input increment leads to the bounded output increment. 
Given the energy of the system, if changes in the control input are limited, changes in the output are also limited and cannot increase indefinitely. $\hfill \hfill \square $
\end{remark}

\begin{lemma} (\rm{\cite{NonlinearS_DD_01}}) \label{le_1}
If the follower systems in \eqref{eq_sys} satisfies Assumptions \ref{as_2}, \ref{as_3} and $|\Delta u_i(k)|\neq 0$ for $\forall k$, then the models of followers can be transformed into
\begin{equation}
    \Delta y_i(k+1)=\phi_i(k)\Delta u_i(k),
\end{equation}
where $\phi_i(k)$ is a pseudo-partial-derivative (PPD) parameter that fulfills $|\phi_i(k)|\leq b_c$.
\end{lemma}

\begin{assumption}\label{as_4}
The sign of the PPD parameter remains unchanged for all $k$ and satisfies $\phi_i(k)>\varepsilon>0$ or $\phi_i(k)<-\varepsilon$. Keeping the generality intact, $\phi_i(k)>\varepsilon$ is assumed in the following discussion. 
\end{assumption}

\begin{remark}
  Most of model-based control methods have a similar assumption as Assumption \ref{as_4}, which means that the input of system increase does not lead to the output decrease. This assumption is crucial to ensure that the movement of systems is in the desired direction even when the systems is nonlinear or even unknown during our tracking process. $\hfill \hfill \square $
\end{remark}

Next, the follower systems in \eqref{eq_sys} could is represented as the following model-free data models according to Lemma \ref{le_1}:
\begin{equation}\label{eq_sys_3}
    y_i(k+1)=y_i(k)+\phi_i(k)\Delta u_i(k),\quad i=1,2,\ldots,N.
\end{equation}

\section{ Problem Formulation  } \label{section3}
In this section, a leader-following consensus of unknown
nonlinear MASs against multiple cyber attacks is considered. The cyber attacks that the systems may suffer during the control process will also be illustrated.

\subsection{Attack Descriptions}\label{attack_descriptions}
Potential attackers who can derive the MASs by launching DoS attacks and AAs are considered in this paper. The specific definitions of the above two attacks will be presented below.

\emph{1) DoS Attacks:} DoS attacks are conducted to derive the control performance of MAS by cutting off the communication channel between agents. Due to the limitation of energy, DoS attacks occur intermittently. The $i$th DoS attacks interval is denoted as
$[T_i^{on},T_i^{off})$, Wherein $T_i^{on}\in \mathbb{N}$ and $T_i^{off}\in \mathbb{N}$ are denoted as the start and end time instant of the DoS attacks in entire control period. The union of DoS attacks in the interval $[0,k]$ with $k\in \mathbb{N}$ can be obtained:
\begin{equation}
  \Xi_d(0,k)=\{\cup_{i\in \mathbb{N}}[T_i^{on},T_i^{off})\}\cap [0,k].
\end{equation}

Next, the union of interval without DoS attack be obtained:
\begin{equation}
  \Xi_s(0,k)=[0,k] \backslash \Xi_d(0,k).
  \end{equation}

\begin{assumption}\label{as_5}
$|\Xi_d(0,k)|$ and $|\Xi_s(0,k)|$ defined as total time interval of DoS Attack and  total time interval without DoS respectively satisfies the following condition:
\begin{subequations}
\begin{flalign}
  |\Xi_a(0,k)|&\leq M+\beta k,\\
  |\Xi_a(0,k)|&+|\Xi_s(0,k)|=k,
\end{flalign}
\end{subequations}
where $M> 0$ and $0< \beta <1$ are constants to be confirmed.
\end{assumption}

\begin{remark}
  Limited by energy, DoS attacks cannot last forever, and there will be certain constraints as Assumption \ref{as_5}. From the perspective of energy, $M$ represents the maximum duration of DoS attacks, which depends on the attacker's own energy storage. $\beta$ represents the charging rate and cannot be greater than $1$. It is assumed that the energy consumption and energy supplement during DoS attack occurring is simultaneous. When $\beta>1$, the energy supplement is greater than the energy consumption, meaning that DoS attack can last forever, which is inconsistent with reality. $\hfill \hfill \square $
\end{remark}

A flag signal introduced to indicate DoS Attack occurs or not is as follows:
\begin{equation*}
\psi(k)=
\begin{cases}
0,& \text{if $k\in \Xi_d(0,k)$},\\
1,& \text{if $k\in \Xi_s(0,k)$}.\\
\end{cases}
\end{equation*}

\emph{2) Unbounded Actuation Attacks:} Actuation attacks from potential attackers acts on the control input of system, by injecting a wrong attack signal into the motor input of each agent to deteriorate system performance. When the MAS is under AAs, the control input of each agent system is shown as follows:
\begin{equation}
    \bar{u}_i(k)=u_i(k)+\chi_i(k),
\end{equation}
in which $\chi_i(k)$ is denoted as the unknown  and possibly unbounded actuation attack signals, and $\bar{u}_i(k)$ is control input of actual actuator being polluted by AAs. 
Although the attack signal can be unbounded, the control input of the system will be limited by the objective conditions. Meanwhile, the unbounded AAs in this paper must meet the following assumption.

\begin{assumption}\label{as_6}
AAs signals grow from zero and the variation of AA signals at each sampling time is bounded by $\bar{d}$, that is, $\chi_i(0)=0$ and $|\Delta\chi_i(k)|<\bar{d}$.
\end{assumption}

\begin{remark}
  Maybe the unbounded Actuation attacks are a bit unrealistic from the perspective of the objective structure and energy limitation of the actuator. However, the unbounded AAs here are only an attack signal passed to the actuator and do not represent the actual offset of the actual actuator input. When the attack signal tends to be unbounded, although the actuator cannot reach infinity due to its structure and energy limitations, it will reach its own input threshold. When designing attack compensation for AAs, we still have to treat attack signals as possibly unbounded.  $\hfill \hfill \square $
\end{remark}

 \subsection{Problem Formulation}
A global tracking error is defined to measure the tracking performance, which is as follows:
\begin{equation}\label{eq_eror_1}
    e_i(k)=y_0(k)-y_i(k),i=1,2,\ldots,N.
\end{equation}

Without being attacked, the local tracking error of the $i$th followers is denoted as
\begin{equation}\label{eq_eror_2}
    \xi_i(k)=\sum_{j\in N_i}a_{i j}(y_j(k)-y_i(k))+c_i(y_0(k)-y_i(k)),
\end{equation}
where $a_{i j}$ are the parameters in the adjacency matrix and $c_i = 1$ or $0$ denotes that if there is a communication channel from leader $0$ to follower $i$ or not.

Based on the settings in Subsection \ref{attack_descriptions}, the nonlinear dynamics of followers in \eqref{eq_sys} affected by AAs are expressed as
\begin{equation}
      \bar{y}_i(k+1)=f_i(\bar{y}_i(k),\bar{u}_i(k)),\quad i=1,2,\ldots,N+1,
\end{equation}
where $\bar{y}_i(k)$ is denoted as state output under cyber attacks.

Considering the damage of both DoS attacks and AAs, the local tracking error in \eqref{eq_eror_2} is deteriorated into, due to the several cyber attacks, the form as
\begin{equation}
    \bar{\xi}_i(k)=\sum_{j\in N_i}a_{i j}^{\psi(k)}(\bar{y}_j(k)-\bar{y}_i(k))+c_i^{\psi(k)}(y_0(k)-\bar{y}_i(k)).
\end{equation}

From the two forms of local tracking error \eqref{eq_eror_1} and \eqref{eq_eror_2}, it is easy to see that the cyber attacks on the system are incredibly destructive. Leader-following consensus control against these attacks will be challenging.

Based on the above discussion about both several attack descriptions and CFDL data models of followers, we will study the leader-following consensus control of MASs against cyber attacks, including DoS attacks and unbounded AAs. The details are as follows.

\noindent \textbf{Problem LFCCA}
 (Leader-following consensus control against composite attacks) : In the case of two malicious cyber attacks described in Subsection \ref{attack_descriptions}, design a novel distributed protocols for systems \ref{eq_sys} based on MFAC so that the global tracking error $e_i(k)$ in \eqref{eq_eror_1} is UUB convergence under the above Assumptions \ref{as_1}-\ref{as_6}, that is, $\lim_{k\to\infty}\|e_i(k)\|\leq B, i=1,2,\ldots, N$.

\section{ Main Results } \label{gl}
Inspired by the recent sprung-up digital twin technology \cite{DT_01}, a double-layer distributed model free adaptive control (DMFAC) framework is investigated in this section. The hierarchal control scheme solves the LFCCA by employing a nonlinear TL against DoS attacks and a distributed adaptive control with attack compensation against AAs on the CPL, both only using the I/O data of MASs.

\subsection{Design of TL  based on Data-driven against Frequency-constrained DoS Attacks}
\begin{figure} 
    \centering
    \includegraphics[width=8cm]{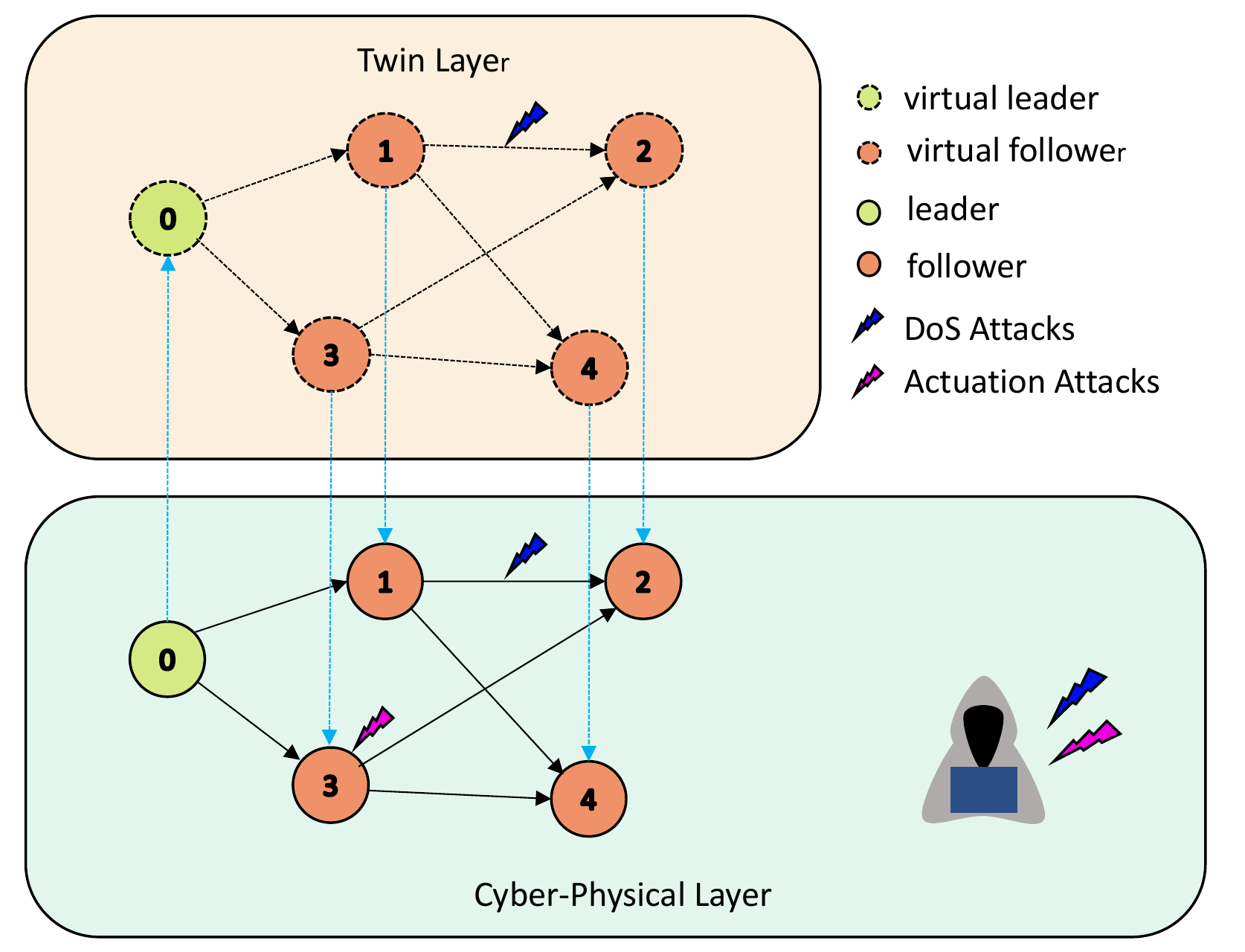}
    \caption{MASs against cyber attacks: A double-layer framework.}\label{TL_CPL}
\end{figure}
In this paper, the MASs is subject to DoS and AAs attacks from covert attackers.
As shown in Fig. \ref{TL_CPL}, a double layer-framework based on the TL is built to  divide the resilient control scheme against cyber attacks into leader-following control against DoS attacks on the TL and point-to-point following against AAs on the CPL.

The TL has superior information privacy and transmits less significant physical signals, making it impervious to many assaults, including AAs. DoS attacks, however, could still work on the TL and cut off the information channel between agents since it can be easy to implement even for attacks without the knowledge of MASs to paralyze the consensusability of MASs with a limited budget as \cite{DC_DoS_01}. 

In the double-layer framework, the leader can transmit self-state information $y_i$ to the TL in real-time, and transmit its state information to the corresponding virtual followers on the TL according to the topological network, which is equivalent to that there is a virtual leader on the TL, and its dynamics and topology are consistent with the actual leader described in \eqref{eq_sys}.

The system models of virtual followers are unknown and nonlinear and reconstructed on the TL as follows:
\begin{equation}\label{eq_vsy}
\begin{cases}
\tilde{y}_0(k+1)=\tilde{f}_0(\tilde{y}_i(k)),\\
\tilde{y}_i(k+1)=\tilde{f}_i(\tilde{y}_i(k),\tilde{u}_i(k)),\quad i=1,2,\ldots,N,
\end{cases}
\end{equation}
where $\tilde{y}_i(k)\in \mathbb{R}$ is the state output, $\tilde{u}_i(k)\in \mathbb{R}$ is control input and $\tilde{f}_i(\cdot)$ is an potentially unknown nonlinear function on the TL, respectively.
\begin{remark}
State information $y_0$ transmitted by actual leader on the CPL to TL is regarded as the state information $\tilde{y}_0$ from virtual leaders on the TL, thus $y_0$ and $\tilde{y}_0$ have the same dynamics, control input and state output. For the convenience of later analysis, $y_0$ is used to express $y_0$ and $\tilde{y}_0$ uniformly. $\hfill \hfill \square $
\end{remark}

In order to improve security and privacy on the TL, the follower models on the TL could be different from that on the CPL, that is, $\tilde{f}_i(\cdot)$ of followers above can be designed as nonlinear uncertainty.

\begin{remark}
The virtual followers on the TL are just some data, and its dynamics can be designed into any ideal form, such as homogeneous or heterogeneous, linear or nonlinear and  known or uncertain. In order to achieve a good control effect, the dynamics of virtual followers can be designed as a simple homogeneous linear form, which also increases the risk of being attacked. Therefore, system dynamics are also designed as unknown nonlinear form like that of followers on the CPL, so as to improve the invisibility of virtual leaders on the TL under attacks. $\hfill \hfill \square $
\end{remark}

The following necessary assumptions are put forward for  completing the leader-following consensus analysis on the TL.
\begin{assumption}\label{as_7}
The partial derivative of $\tilde{f}_i(\cdot)$ with respect to $\tilde{u}_i(k)$ is continuous.
The sytems of virtual follower in \eqref{eq_vsy} meets the generalized Lipschitz condition, that is, if $\Delta \tilde{u}_i(k)\neq 0$, $\Delta \tilde{y}_i(k+1)|\leq b_t \Delta \tilde{u}_i(k)|$.
if $|\Delta \tilde{u}_i|\neq 0$ for $\forall k$, then the system can be transformed into $\Delta \tilde{y}(k+1)=\Phi(k)\Delta \tilde{u}(k)$, where $\Phi(k)$ is bounded and satisfies $0<\varepsilon<\Phi(k)\leq b_t$.
\end{assumption}

\begin{remark}
  Assumption \ref{as_7} for virtual followers on the TL is similar to above Assumption \ref{as_2} and \ref{as_3} for followers on the CPL, all of that is necessary assumptions to convert the unknown nonlinear system models into  data-driven models. $\hfill \hfill \square $
\end{remark}

Next, a data-driven model of virtual followers similar to \eqref{eq_sys_3} is obtained as
\begin{equation}
        \tilde{y}_i(k+1)=\tilde{y}_i(k)+\Phi_i(k)\Delta \tilde{u}_i(k),\quad i=1,2,\ldots,N.
\end{equation}

Due to the difficulty in obtaining the value of the PPD parameter, the estimator is required to estimate $\Phi(k)$ as follows:
\begin{align}\label{eq_16}
 \hat{\Phi}_i(k)=&\hat{\Phi}_i(k-1)+\frac{\eta_t \Delta \tilde{u}_i(k-1)}{\mu_t +{\Delta \tilde{u}_i(k-1)}^2}\nonumber\\ 
&\times [\Delta \tilde{y}_i(k)-\hat{\Phi}_i(k-1) \Delta \tilde{u}_i(k-1)],
\end{align}

where $\eta_t\leq 1$ is a step coefficient and $\mu_t$ is positive constant as penalty factor.
\begin{remark}
  The estimation algorithm \eqref{eq_16} of PPD parameter is obtained by minimizing the performance function as follows:
  \begin{equation*}
  \begin{aligned}      
      J_1[\hat{\Phi}_i(k)]=&{[\Delta \tilde{y}_i(k)-\hat{\Phi}_i(k) \Delta \tilde{u}_i(k-1)]}^2\\
      &+\mu_t{[\hat{\Phi}_i(k)-\hat{\Phi}_i(k-1)]}^2.     
\end{aligned}
  \end{equation*}
$\hfill\hfill\square$
\end{remark}

Define the following local tracking errors of virtual follower $i$ on the TL:
\begin{equation}\label{eq_E_TL}
\begin{aligned}
    \tilde{\xi}_i(k)&=\sum_{j\in N_i}a_{i j}(\tilde{y}_j(k)-\tilde{y}_i(k))+c_i(y_0(k)-\tilde{y}_i(k))\\
    &=\sum_{j\in N_i}a_{i j}(\tilde{e}_i(k)-\tilde{e}_j(k))+c_i \tilde{e}_i(k),
\end{aligned}
\end{equation}
where $\tilde{e}_i(k)=y_0(k)-\tilde{y}_i(k)$ is global errors on the TL, $a_{i j}$ is the parameters in the adjacency matrix and $c_i = 1$ or $0$ denotes that there is a communication channel from leader $0$ to agent $i$ or not.

Then we obtain the control algorithm:
\begin{equation}\label{eq_control_u1}
\tilde{u}_i(k)=\tilde{u}_i(k-1)+\frac{\gamma_t \hat{\Phi}_i(k)}{\lambda_t +{\hat{\Phi}_i(k)}^2} \tilde{\xi}_i(k),
\end{equation}
where $\gamma_t <1$ is a step coefficient and $\lambda_t$ is positive constant as penalty factor.
\begin{remark}
  Similar to the estimation algorithm above, the control law is obtained according to following performance function:
\begin{equation*}
  J_2[\tilde{u}_i(k)]={\tilde{\xi}_i(k+1)}^2+\lambda_t{[\tilde{u}_i(k)-\tilde{u}_i(k-1)]}^2. 
  \end{equation*}
$\hfill\hfill\square$
\end{remark}

Considering the DoS attacks, flag signal $\psi(k)$ is introduced into the control law, that is,
\begin{equation}\label{eq_control_u2}
    \tilde{u}_i(k)=\tilde{u}_i(k-1)+\psi(k)\frac{\gamma_t \hat{\Phi}_i(k)}{\lambda_t +{\hat{\Phi}_i(k)}^2} \Tilde{\xi}_i(k).
\end{equation}
\begin{remark}
  When DoS attack occurs, the communication between virtual agents is interrupted, so that the followers cannot get the state information of the virtual leader and that of neighbors' agents to update the control input according to control algorithm \eqref{eq_control_u1}. 
  As a result, a defense strategy against DoS described by \eqref{eq_control_u2} is designed, which only uses state information before the attack. We took the conservative defensive strategy to keep the control input unchanged, enabling followers to maintain the original motion state under DoS attacks and to follow the virtual leader after DoS attacks.  $\hfill \hfill \square $
\end{remark}

Thus, a novel DMFAC algorithm based on data-driven against DoS attacks on the TL is proposed as
\begin{align}\label{eq_control_DoS}
   \hat{\Phi}_i(k)=&\hat{\Phi}_i(k-1)
   +\frac{\eta_t \Delta \tilde{u}_i(k-1)}{\mu_t +{\Delta \tilde{u}_i(k-1)}^2}\nonumber\\
   &\times[\Delta \tilde{y}_i(k)-\hat{\Phi}_i(k-1) \Delta \tilde{u}_i(k-1)],\nonumber\\
  \hat{\Phi}_i(k)=&\hat{\Phi}_i(0), \text{   if $|\hat{\Phi}_i(k)| < \varepsilon$ or ${\rm sign}(\hat{\Phi}_i(k))={\rm sign}(\hat{\Phi}_i(1))$}, \nonumber\\
\tilde{u}_i(k)=&\tilde{u}_i(k-1)+\psi(k)\frac{\gamma_t \hat{\Phi}_i(k)}{\lambda_t +{\hat{\Phi}_i(k)}^2} \Tilde{\xi}_i(k).
\end{align}

\begin{myTheo}\label{Th_1}
Considering the MASs mentioned at \eqref{eq_vsy} satisfies Assumptions \ref{as_7}, leader-following consensus on the TL is achieved by \eqref{eq_control_DoS} if following conditions are satisfied, so that global tracking errors $\tilde{e}_i$ is UUB  convergence. 
\begin{subequations}
\begin{flalign}
    \frac{b_t\gamma_t\max_{i\in N}(\sum_{j\in N_i} a_{i j}+c_i)}{2\sqrt{\lambda_t}}<1,\\
    \beta<\frac{-\ln{\alpha_1}}{\ln{\alpha_2}-\ln{\alpha_1}}<1.
\end{flalign}
\end{subequations}
\end{myTheo}

\textbf{Proof.} 
Let
\begin{equation*}
\begin{aligned}
    \tilde{\xi}(k)
=&
 \left[
  \begin{array}{ccc}
 \tilde{\xi}_1(k)\\
\tilde{\xi}_2(k)\\
\vdots\\
\tilde{\xi}_N(k)\\
  \end{array}
  \right],~\tilde{E}(k)
  =
  \left[
  \begin{array}{ccc}
 \tilde{e}_1(k)\\
\tilde{e}_2(k)\\
\vdots\\
\tilde{e}_N(k)\\
  \end{array}
  \right],\\
  ~\tilde{U}(k)
  =&
  \left[
  \begin{array}{ccc}
 \tilde{u}_1(k)\\
\tilde{u}_2(k)\\
\vdots\\
\tilde{u}_N(k)\\
  \end{array}
  \right].
\end{aligned}
\end{equation*}
The compact representation of local tracking errors in \eqref{eq_E_TL} is defined as
\begin{equation}
    \tilde{\xi}(k)=(L+C)\tilde{E}(k).
\end{equation}
Next, the compact form of control law \eqref{eq_control_u2} could be obtained as follow:
\begin{equation}
    \tilde{U}(k)=\tilde{U}(k-1)+\psi(k)P(k)(L+C)\tilde{E}(k),
\end{equation}
where $P(k)={\rm diag}(\rho_1(k),\rho_2(k),\ldots,\rho_N(k))$ and $\rho_i(k)=\frac{\gamma_t \hat{\Phi}_i(k)}{\lambda_t +{
\hat{\Phi}_i(k)}^2}$ and 
matrix $L$ and 
matrix $C$ is defined in \ref{graph}.

When DoS Attack launches, that is, $k\in \Xi_s(0,k)$, the leader-following tracking error on the TL can be obtained:
\begin{equation}
\begin{aligned}
\tilde{E}(k+1)
&=Y_0(k+1)-\tilde{Y}(k+1)\\
&=Y_0(k)-\tilde{Y}(k)-\Delta \tilde{Y}(k+1) +\Delta Y_0(k+1)\\
&=\tilde{E}(k)-\Phi(k)\Delta \tilde{U}(k)+\Delta Y_0(k+1)\\
&=(I_N-\Phi(k) P(k)(L+C))\tilde{E}(k)+\Delta Y_0(k+1)\\
&=(I_N-G(k))\tilde{E}(k)+\Delta Y_0(k+1),
\end{aligned}
\end{equation}
where $\Phi(k)=diag(\Phi_1(k),\Phi_2(k),\cdots,\Phi_N(k)), \\\tilde{Y}(k)={[\tilde{y}_1(k),\tilde{y}_2(k),\cdots,\tilde{y}_N(k)]}^{\mathrm{T}}, \\\Delta{\tilde{Y}}(k)={[\Delta{\tilde{Y}}_1(k),\Delta{\tilde{Y}}_2(k),\cdots,\Delta{\tilde{Y}}_N(k)]}^{\mathrm{T}}, \\Y_0(k)={[y_0(k),y_0(k),\cdots,y_0(k)]}^{\mathrm{T}}$ \\and ${\Delta{Y}_0(k)={[\Delta{y}_0(k),\Delta{y}_0(k),\cdots,\Delta{y}_0(k)]}^{\mathrm{T}}}$\\are N-dimensional column vectors.

Since $0<\Phi(k)\leq \bar{\Phi}=b_t,0<\rho_i(k)\leq \bar{\rho}=\frac{\gamma_t}{2\sqrt{\lambda_t}}$ and $0<\frac{\gamma_t \hat{\Phi}_i(k)}{\lambda_t +{\hat{\Phi}_i(k)}^2}\leq \frac{\gamma_t \hat{\Phi}_i(k)}{2 \sqrt{\lambda_t}|\hat{\Phi}_i(k)|}=\frac{\gamma_t}{2\sqrt{\lambda_t}}$,
we have
\begin{equation}
    0<\|G(k)\|\leq
    \overline{G}=\frac{b_t\gamma_t\max_{i\in N}(\sum_{j\in N_i} a_{i j}+c_i)}{2\sqrt{\lambda_t}}.
\end{equation}
Let 
\begin{equation}
    \frac{b_t\gamma_t\max_{i\in N}(\sum_{j\in N_i} a_{i j}+c_i)}{2\sqrt{\lambda_t}}<1,
\end{equation}
meaning that $ 0<\|G(k)\|<1$, so that the matrix $[I_N-G(k)]$ is an irreducible sub-stochastic matrix. Next, a maximum error is defined as $\tilde{e}_{\max}(k)=\max(\tilde{e}_1(k),\tilde{e}_2(k),\ldots,\tilde{e}_N(k))$ and $\|\Delta Y_0(k+1)\|\leq \Omega$ is assumed, we have
\begin{equation}
    |\tilde{e}_{\max}(k+1)|\leq \alpha_1|\tilde{e}_{\max}(k)|+\Omega,
\end{equation}
where $\alpha_1=1-\underline{G}<1$ and $\underline{G}$ is the minimum value of $G(k)$. 
\begin{remark}
  Because of the lag of communication, we can only get the current control input based on the current information, and the most ideal tracking effect also exist tracking errors due to the change of the leader's state $\|\Delta Y_0(k+1)\|$. So the upper bound of $\|\Delta Y_0(k+1)\|$ is introduced here to constrain the upper bound of global tracking error $\tilde{e}_i$.   $\hfill \hfill \square $
\end{remark}

When the system encounters DoS attacks, that is, $k\in \Xi_a(0,k)$, flag signal $\psi(k)$ defined in \ref{attack_descriptions} is zeros, that is, $\psi(k)=0$. In this situation, $\tilde{u}_i(k)=\tilde{u}_i(k-1)$ in line with \eqref{eq_control_u2}. According to Lemma \ref{le_1}, one gets
\begin{equation}
    \tilde{Y}(k+1)=\tilde{Y}(k).
\end{equation}
Then the tracking error can be rewritten as follows:
\begin{equation}
\begin{aligned}
\tilde{E}(k+1)
&=Y_0(k+1)-\tilde{Y}(k+1)\\
&=Y_0(k)-\tilde{Y}(k)+\Delta \tilde{Y}_0(k+1)\\
&=\tilde{E}(k)+\Delta Y_0(k+1).\\
\end{aligned}
\end{equation}

Next, we have
\begin{equation}
    |\tilde{e}_{\max}(k+1)|\leq \alpha_2|\tilde{e}_{\max}(k)|+\Omega,
\end{equation}
where $\alpha_2>1$.

In summary, the max global tracking error on the TL
is represented as 
\begin{equation}
|\tilde{e}_{\max}(k+1)|=
\begin{cases}
\leq \alpha_1|\tilde{e}_{\max}(k)|+\Omega,& \text{if $k\in \Xi_s(0,k)$},\\
\leq \alpha_2|\tilde{e}_{\max}(k)|+\Omega,& \text{if $k\in \Xi_a(0,k)$}.\\
\end{cases}
\end{equation}
 
It should be noted that MASs on the TL will be in two situations according to the difference of $k$, that is, $k\in \Xi_s(0,k)$ and $k\in \Xi_a(0,k)$. Next, one case as $k\in \Xi_a(0,k)$ is discussed as follows, and the discussion of the second case is similar to the first case.
    \begin{align}\label{eq_emax}
     &|\tilde{e}_{\max}(k+1)|\nonumber\\
    &<\alpha_2|\tilde{e}_{\max}(k)|+\Omega\nonumber\\
    &\leq\alpha_2^2|\tilde{e}_{\max}(k-1)|+\alpha_2\Omega+\Omega\nonumber\\
    &\leq\alpha_2^{k-T_i^{on}+1}|\tilde{e}_{\max}(T_i^{on})|+\sum_{n=0}^{k-T_i^{on}}\alpha_2^n\Omega\nonumber\\
    &\leq\alpha_2^{k-T_i^{on}+1}(\alpha_1|\tilde{e}_{\max}(T_i^{on}-1)|+\Omega)\nonumber\\
    &\quad +\sum_{n=0}^{k-T_i^{on}}\alpha_2^n\Omega\nonumber\\
    &\leq\alpha_2^{k-T_i^{on}+1}(\alpha_1^2|\tilde{e}_{\max}(T_i^{on}-2)|+\alpha_1\Omega+\Omega)\nonumber\\
    &\quad+\sum_{n=0}^{k-T_i^{on}}\alpha_2^n\Omega\nonumber\\
    &\leq\alpha_2^{k-T_i^{on}+1}(\alpha_1^{T_i^{on}-T_{i-1}^{off}}|\tilde{e}_{\max}(T_{i-1}^{off})|\nonumber\\
    &\quad+\sum_{n=0}^{T_i^{on}-T_{i-1}^{off}-1}\alpha_1^n\Omega)+\sum_{n=0}^{k-T_i^{on}}\alpha_2^n\Omega\nonumber\\
    &\leq\alpha_2^{k-T_i^{on}+1}\alpha_1^{T_i^{on}-T_{i-1}^{off}}|\tilde{e}_{\max}(T_{i-1}^{off})|\nonumber\\
    &\quad+\alpha_2^{k-T_i^{on}+1}\sum_{n=0}^{T_i^{on}-T_{i-1}^{off}-1} \alpha_1^{n}\Omega+\sum_{n=0}^{k-T_i^{on}}\alpha_2^n\Omega.
    \end{align}

when $|\Xi_s(0,k)|=k-|\Xi_a(0,k)|$ and $|\Xi_a(0,k)|\leq M+\beta k$ mentioned at Assumption \ref{as_5} is introduced in \eqref{eq_emax}, the local tracking error  is transformed into

\begin{align}
&|\tilde{e}_{\max}(k+1)|\nonumber\\
&\leq\alpha_1^{k-|\Xi_a(0,k)|}\alpha_2^{|\Xi_a(0,k)|}|\tilde{e}_{\max}(0)|\nonumber\\
&\quad+\sum_{j=0}^k \alpha_1^{k-j-|\Xi_a(j,k)|}\alpha_2^{|\Xi_a(j,k)|}\Omega\nonumber\\
&\leq e^{(k-|\Xi_a(0,k)|)\ln{\alpha_1}+|\Xi_a(0,k)|\ln{\alpha_2}}|\tilde{e}_{\max}(0)|\nonumber\\
&\quad+\sum_{j=0}^k e^{(k-j-|\Xi_a(j,k)|)\ln{\alpha_1}+|\Xi_a(j,k)|\ln{\alpha_2}}\Omega\nonumber\\
&\leq e^{(k-M-\beta k)\ln{\alpha_1}+(M+\beta k)\ln{\alpha_2}}|\tilde{e}_{\max}(0)|\nonumber\\
&\quad+\sum_{j=0}^k e^{(k-j-M-\beta (k-j))\ln{\alpha_1}+(M+\beta(k-j))\ln{\alpha_2}}\Omega\nonumber\\
&\leq e^{(\ln{\alpha_2}-\ln{\alpha_1})M+[\ln{\alpha_1}+\beta(\ln{\alpha_2}-\ln{\alpha_1})]k}|\tilde{e}_{\max}(0)|\nonumber\\
&\quad+\sum_{j=0}^k e^{(\ln{\alpha_2}-\ln{\alpha_1})M+[\ln{a_1}+\beta(\ln{\alpha_2}-\ln{\alpha_1})](k-j)}\Omega\nonumber\\
&\leq e^{(\ln{\alpha_2}-\ln{\alpha_1})M}e^{[\ln{\alpha_1}+\beta(\ln{\alpha_2}-\ln{\alpha_1})]k}|\tilde{e}_{\max}(0)|\nonumber\\
&\quad+e^{(\ln{\alpha_2}-\ln{\alpha_1})M}\Omega\nonumber\\
&\quad\times \sum_{j=0}^k e^{[\ln{\alpha_1}+\beta(\ln{\alpha_2}-\ln{\alpha_1})](k-j)}.
\end{align}

Let $\ln{\alpha_1}+\beta(\ln{\alpha_2}-\ln{\alpha_1})<0$, the maximum tracking error $\tilde{e}_{\max}$ is uniformly bounded, that is, global tracking error $\tilde{e}_i$ on the TL is UUB with error upper $B_t$ as follows:
\begin{equation}
    |\tilde{e}_{\max}(k+1)|\leq \frac{e^{(\ln{\alpha_2}-\ln{\alpha_1})M}}{1-e^{[\ln{\alpha_1}+\beta(\ln{\alpha_2}-\ln{\alpha_1})]}}\Omega=B_t.
\end{equation}

The proof is completed.  $\hfill \hfill \blacksquare $

\subsection{One-to-One Tracking between CPL and TL against Unbounded AAs}

Tracking errors between TL and CPL is defined as
\begin{equation}
    \sigma_i(k)=\tilde{y}_i(k)-y_i(k)
\end{equation}

In Theorem \ref{Th_1}, we have proved that TL can resist DoS attacks with limited attack frequency, and the global tracking error on the TL is UUB convergence. In the following, We prove that the global tracking error on the CPL is also UUB convergence under unbounded AAs. 

According to the general data-driven method, a 
DMFAC is proposed to drive $y_i(k)$ to $\tilde{y}_i(k)$ without considering AAs described as

\begin{align}\label{eq_36}
   \hat{\phi}_i(k)=&\hat{\phi}_i(k-1)+\frac{\eta_c \Delta u_i(k-1)}{\mu_c +{\Delta u_i(k-1)}^2}\nonumber\\
   &\times[\Delta y_i(k)-\hat{\phi}_i(k-1) \Delta u_i(k-1)],\nonumber\\
  \hat{\phi}_i(k)=&\hat{\phi}_i(0), \text{   if $|\hat{\phi}_i(k)| < \varepsilon$ or ${\rm sign}(\hat{\phi}_i(k))={\rm sign}(\hat{\phi}_i(0))$},\nonumber\\
u_i(k)=&u_i(k-1)+\frac{\gamma_c \hat{\phi}_i(k)}{\lambda_c +{\hat{\phi}_i(k)}^2}[\tilde{y}_i(k+1)-y_i(k)],
\end{align}

where $\lambda_c$ and $\mu_c$ are
positive penalty factor and $0<\eta_c<1$ and $0<\gamma_c<1$ are step coeﬀicient.
\begin{remark}
Generally speaking, we can't get the value of $\tilde{y}$ at $k+1$ ahead of current time $k$. However, $\tilde{y}$ is only the result of data operation in the digital twin layer, so the value of $\tilde{y}$ at $k+1$ can be calculated immediately according to the state and input at time $k$, and transmitted to the agents in the CPL through the channel between the CPL and TL. This is why $\tilde{y}(k+1)$ is used in \eqref{eq_36} at current time $k$. $\hfill \hfill \square $
\end{remark}

Actuation attacks signal is a bias signal injected into the control input, deteriorating the tracking performance of MASs. When encountering unbounded AAs, the control input of each followers is actually described as:
\begin{align}
    \bar{u}_i(k)&=u_i(k-1)+\frac{\gamma_c \hat{\phi}_i(k)}{\lambda_c +{\hat{\phi}_i(k)}^2}[\tilde{y}_i(k+1)-y_i(k)]+\chi_i(k)\nonumber\\
    &=\bar{u}_i(k-1)+\frac{\gamma_c \hat{\phi}_i(k)}{\lambda_c +{\hat{\phi}_i(k)}^2}[\tilde{y}_i(k+1)-y_i(k)]+\Delta \chi_i(k),
\end{align}

where $\chi_i(k)$ is actuation attack signals and $\Delta \chi_i(k)=\chi_i(k)-\chi_i(k-1)$.

In the data-driven model presented in \eqref{eq_sys_3}, $\Delta u_i(k)$ is used as the system input instead of $u_i(k)$. Therefore, when analyzing the impact of actuation attack signal $\chi_i(k)$ on system stability, analyzing $\Delta \chi_i(k)$ is necessary.
The value of $\Delta\chi_i(k)$ is unknown and can not be obtained. Therefore, a estimator is designed to estimate the value of  $\Delta \chi_i(k)$ as follows:
\begin{equation} \label{eq_control_compensattion_AAs}
    \Delta \hat{\chi}_i(k)=
\begin{cases}
\frac{\bar{d}[\Delta \hat{\chi}_i(k-1)-r_i(k)\sigma_i(k)]}{\bar{d}+|\Delta \hat{\chi}_i(k-1)-r_i(k)\sigma_i(k)|}& \text{if $k>0$}\\
0& \text{if $k=0$},\\
\end{cases}
\end{equation}
where $\bar{d}$ is the upper variation of actuation attack signals defined in Assumption \ref{as_6}, $\Delta \hat{\chi}_i(k)$ is the estimated value of $\Delta \chi_i(k)$ and $r_i(k)=\frac{\gamma \hat{\phi}_i(k)}{\lambda_c +{|\hat{\phi}_i(k)|}^2}$ is a adaptive scale factor with $0<\gamma <\gamma_c$.
\begin{remark}
  The attack signal $\chi_i$ acts on the controller input, and finally affects the point-to-point tracking error $\sigma_i$ between CPL and TL. Because there is a causal relationship between $\Delta \chi_i$ and $\sigma_i$, it is an effective and feasible method to reconstruct $\Delta \chi_i$ according to the tracking error $\sigma_i$. $\hfill \hfill \square $
\end{remark}

 Considering the compensation of $\Delta \chi_i(k)$, a novel 
DMFAC is proposed  as

\begin{align}\label{eq_control_AAs}
   \hat{\phi}_i(k)=&\hat{\phi}_i(k-1)+\frac{\eta_c [\Delta u_i(k-1)+\Delta \hat{\chi}_i(k-1)]}{\mu_c +{|\Delta u_i(k-1)+\Delta \hat{\chi}_i(k-1)|}^2}\nonumber\\
   &\times[\Delta y_i(k)-\hat{\phi}_i(k-1) [\Delta u_i(k-1)+\Delta \hat{\chi}_i(k-1)]]\nonumber\\
 \hat{\phi}_i(k)=&\hat{\phi}_i(0), \text{   if $|\hat{\phi}_i(k)| < \varepsilon$ or ${\rm sign}(\hat{\phi}_i(k))={\rm sign}(\hat{\phi}_i(0))$},\nonumber\\
u_i(k)=&u_i(k-1)+\frac{\gamma_c \hat{\phi}_i(k)}{\lambda_c {\hat{\phi}_i(k)}^2}[\tilde{y}_i(k+1)-y_i(k)]-\Delta \hat{\chi}_i(k),
\end{align}

\begin{myTheo} \label{Th_2}
\textbf{Problem LFCCA} is solved by 
double-layer DMFAC algorithm in \eqref{eq_control_DoS}, \eqref{eq_control_compensattion_AAs} and \eqref{eq_control_AAs} if following conditions hold simultaneously:
\begin{subequations}
\begin{flalign}
    &\frac{b_t\gamma_t\max_{i\in N}(\sum_{j\in N_i} a_{i j}+c_i)}{2\sqrt{\lambda_t}}<1,\\
    &\beta<\frac{-\ln{\alpha_1}}{\ln{\alpha_2}-\ln{\alpha_1}}<1,\\
    &0<\frac{\gamma_c b_c}{2\sqrt{\lambda_c}}<1.
\end{flalign}    
\end{subequations}
\end{myTheo}

\begin{figure} 
    \centering
    \includegraphics[width=9cm]{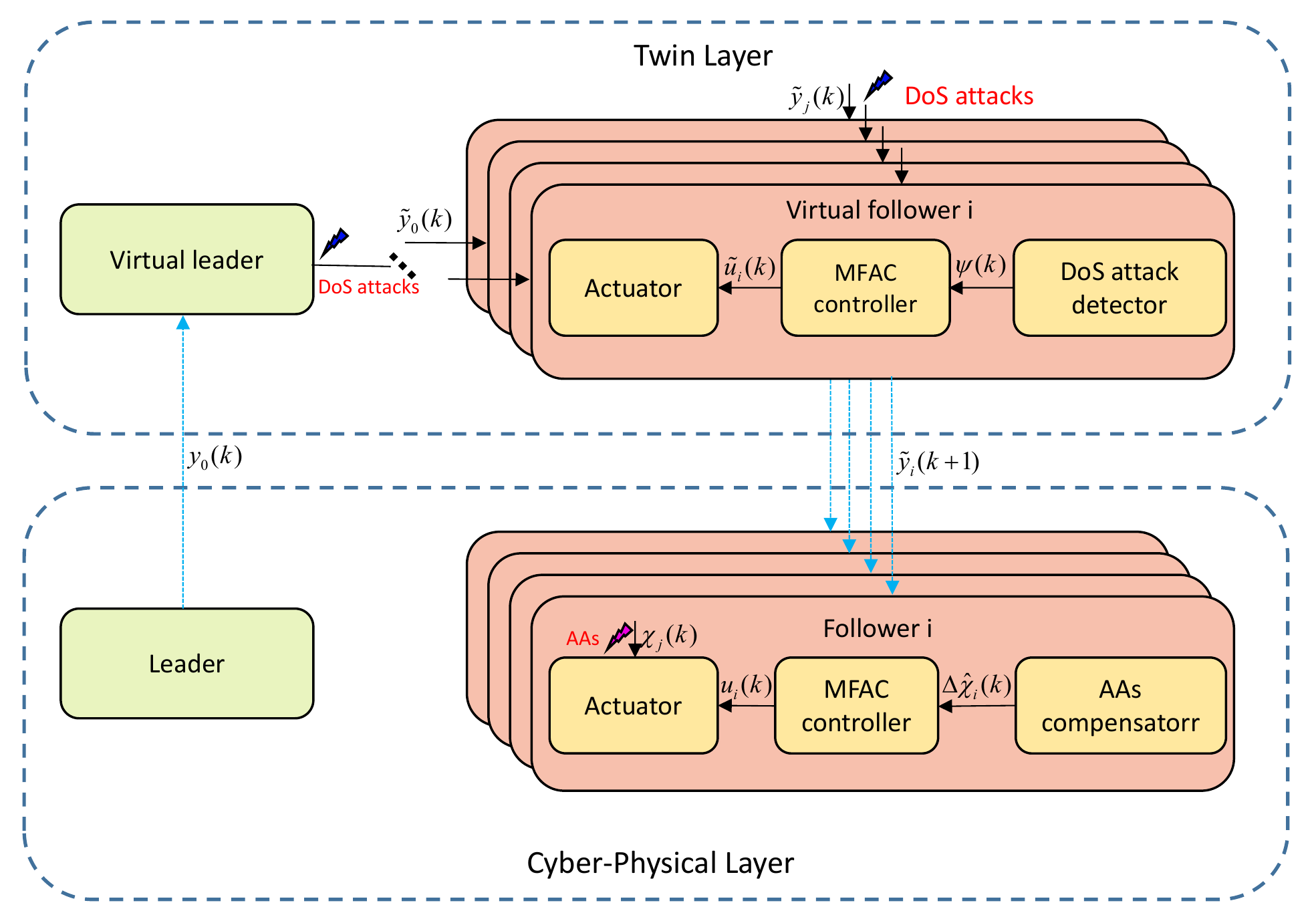}
    \caption{Double-layer DMFAC control framework.}\label{MAFC}
\end{figure}

\textbf{Proof.}
Obviously, $\hat{\phi_i}$ is bounded if the reset mechanism is activated. The situation that the reset mechanism is not activated is discussed below. Let $\tilde{\phi_i}=\hat{\phi_i}-\phi_i$, one gets

\begin{align}
&\tilde{\phi_i}(k)\nonumber\\
&=\hat{\phi_i}(k)-\phi_i(k)\nonumber\\
&=\hat{\phi_i}(k-1)-\phi_i(k)+\frac{\eta_c [\Delta u_i(k-1)+\Delta \hat{\chi}_i(k-1)]}{\mu_c +{|\Delta u_i(k-1)+\Delta \hat{\chi}_i(k-1)|}^2}\nonumber\\
&\quad\times[\Delta y(k)-\hat{\phi_i}(k-1) [\Delta u_i(k-1)+\Delta \hat{\chi}_i(k-1)]]\nonumber\\
&=(1-\frac{\eta_c {[\Delta u_i(k-1)+\Delta \hat{\chi}_i(k-1)]}^2}{\mu_c +{|\Delta u_i(k-1)+\Delta \hat{\chi}_i(k-1)|}^2})\tilde{\phi_i}(k-1)\nonumber\\
&\quad+\phi_i(k-1)-\phi_i(k)+\frac{\eta_c {[\Delta u_i(k-1)+\Delta \hat{\chi}_i(k-1)]}}{\mu_c +{|\Delta u_i(k-1)+\Delta \hat{\chi}_i(k-1)|}^2}\nonumber\\
&\quad\times(\Delta \chi_i(k-1)-\Delta \hat{\chi}_i(k-1))
\end{align}

Since $0<\eta_c<1$, we have $\frac{\eta_c {[\Delta u_i(k-1)+\Delta \hat{\chi}_i(k-1)]}^2}{\mu_c +{|\Delta u_i(k-1)+\Delta \hat{\chi}_i(k-1)|}^2}<1$.  It is easy to prove that $|\frac{\eta_c [\Delta u_i(k-1)+\Delta \hat{\chi}_i(k-1)]}{\mu_c +{|\Delta u_i(k-1)+\Delta \hat{\chi}_i(k-1)|}^2}|<1$ and $\Delta |\hat{\chi}_i(k)|< \bar{d}$. Combined with $\phi_i < b_c$ and $|\Delta \chi_i(k)|<\bar{d}$, one gets
\begin{equation}
    \begin{aligned}
    |\tilde{\phi_i}(k)|\leq& \frac{\eta_c {[\Delta u_i(k-1)+\Delta \hat{\chi}_i(k-1)]}^2}{\mu_c +{|\Delta u_i(k-1)+\Delta \hat{\chi}_i(k-1)|}^2}|\tilde{\phi_i}(k-1)|+2b_c\\
    &+2|\frac{\eta_c [\Delta u_i(k-1)+\Delta \hat{\chi}_i(k-1)]}{\mu_c +{|\Delta u_i(k-1)+\Delta \hat{\chi}_i(k-1)|}^2}|\phi_i(k-1)\bar{d}\\
    \leq&{g}^{k-1}\|\tilde{\phi_i}(1)\|+\frac{2 b_c(1+\bar{d})}{1-g},
    \end{aligned}
\end{equation}
where $0<\frac{\eta_c {[\Delta u_i(k-1)+\Delta \hat{\chi}_i(k-1)]}^2}{\mu_c +{|\Delta u_i(k-1)+\Delta \hat{\chi}_i(k-1)|}^2}<g<1$.

Thus, the Boundedness of $\tilde{\phi}$ is demonstrated. According to the definition of $\tilde{\phi}$ and $\phi<b_c$, it can be obtained that $\hat{\phi}$ is bounded.

So the global tracking errors on the CPL gets
\begin{equation}
    \begin{aligned}
    e_i(k+1)=&y_0(k+1)-y_i(k+1)\\
    =&y_0(k)-y_i(k)+\Delta y_0(k+1)-\phi_i(k)\Delta u_i(k)\\
    =&(1-\frac{\gamma_c \phi_i(k)\hat{\phi}_i(k)}{\lambda_c +{\hat{\phi}_i(k)}^2})e_i(k)+\Delta y_0(k+1)\\
    &+\frac{\gamma_c \phi_i(k)\hat{\phi}_i(k)}{\lambda_c +{\hat{\phi}_i(k)}^2}[y_0(k)-\tilde{y}_i(k+1)]\\
    &-\phi_i(k)\Delta \chi_i(k)+\phi_i(k)\Delta \hat{\chi}_i(k)\\
    =&(1-\frac{\gamma_c \phi_i(k)\hat{\phi}_i(k)}{\lambda_c +{\hat{\phi}_i(k)}^2})[e_i(k)+\Delta y_0(k+1)]\\
    &+\frac{\gamma_c \phi_i(k)\hat{\phi}_i(k)}{\lambda_c +{\hat{\phi}_i(k)}^2}\tilde{e}_i(k+1)-\phi_i(k)\Delta \chi_i(k)\\
    &+\phi_i(k)\Delta \hat{\chi}_i(k)
    \end{aligned}
\end{equation}

According to Theorem \ref{Th_1}, the global tracking error on the TL is UUB convergence, that is, $\tilde{e}_i(k+1)\leq B_t$. Next, the global tracking error on the CPL is Further simplified into

\begin{align}
e_i(k) =&(1-\frac{\gamma_c \phi_i(k)\hat{\phi}_i(k)}{\lambda_c +{\hat{\phi}_i(k)}^2})[e_i(k)+\Delta y_0(k+1)]\nonumber\\
&+\frac{\gamma_c \phi_i(k)\hat{\phi}_i(k)}{\lambda_c +{\hat{\phi}_i(k)}^2}\tilde{e}_i(k+1)-\phi_i(k)\Delta \chi_i(k)\nonumber\\
&+\phi_i(k)\Delta \hat{\chi}_i(k)\nonumber\\
\leq& \alpha e_i(k)+\alpha \Omega +(1-\alpha)B_c+2 b_c \overline{d}\nonumber\\
\leq& \alpha^k e_i(0)+B_t+\frac{2 b_c \overline{d}+\alpha \Omega}{1-\alpha}\nonumber\\
\leq& \alpha^k e_i(0)+B,
\end{align}

in which $|\Delta y_0(k)|<\Omega$, $|\Delta \chi_i(k)|<\overline{d}$,$|\Delta \hat{\chi}_i(k)|<\overline{d}$, $0<\frac{\gamma_c \phi_i(k)\hat{\phi}_i(k)}{\lambda_c +{|\hat{\phi}_i(k)|}^2}\leq\frac{\gamma_c b_c}{2\sqrt{\lambda_c}}<1$ as $\alpha=\max(1-\frac{\gamma_c \phi_i(k)\hat{\phi}_i(k)}{\lambda_c +{|\hat{\phi}_i(k)|}^2})<1$ and $B=B_t+\frac{2 b_c \overline{d}+\alpha \Omega}{1-\alpha}$.

So the global tracking errors $e_i(k)$ on the CPL is UUB convergence. The proof is completed.   $\hfill \hfill \blacksquare $

According to the conclusion of Theorem \ref{Th_2}, a double-layer DMFAC framework based on the TL is proposed to solve the \textbf{Problem LFCCA}. As shown in Fig. \ref{MAFC}, 
leaders on the CPL transmit their state information to virtual leader on the TL. Virtual agents on the TL communicate with each other through topological network, and finally realize MFAC-based leader-following control on the TL. Because there are only some data calculations on digital TL, virtual followers can calculate the state information of the next moment through control input from DMFAC algorithm according to its own system model. Followers on the CPL receive the state information of the virtual follower at the next moment as tracking reference point. The detailed steps of the double-layer DMFAC algorithm are described in Algorithm \ref{alg_1}.

 \begin{algorithm}[t]
 \caption{Double-layer DMFAC Algorithm}\label{alg_1}
 \KwIn{Control input ${u_i(k-1)}$ and change of control input ${\Delta u_i(k-1)}$ at time $k-1$.}
 \KwOut{Current time $k$ control input $u_i(k)$ and change of control input $\Delta u_i(k)$.}

  \textbf{Initialize:} For agent $i$, $i\in \mathcal{V}$, initialize control parameters $\eta_c$, $\mu_c$, $\gamma_c$, $\lambda_c$ and initial value estimated $\hat{\phi_i}(0)$ of PPD parameters of DMFAC on the CPL, parameters $\gamma$ and ${\Delta \chi_i(0)=0}$ of attack compensator for unbounded AAs. Build TL, initialize control parameters $\eta_t$, $\mu_t$, $\gamma_t$, $\lambda_t$ and initial value estimated $\hat{\Phi_i}(0)$ of PPD parameters of DMFAC on the TL, flag signal $\psi(1)=1$ of DoS attacks, $\tilde{u}_i(0)$ and $\Delta\tilde{u}_i(0)$. Let $k=1$, update $\tilde{y}_i(1)$ and $\Delta\tilde{y}_i(1)$ according to $\tilde{u}_i(0)$ and $\Delta\tilde{u}_i(0)$\\
  \For{$k = 1 \to \rm {maxtime}$}{
  \For{$i = 1\to N$}{
  \tcp{TL}
 Receive $\tilde{y}_0(k)$ and update $\psi(k)$. \\
  Update $\hat{\Phi_i}(k)$, $\tilde{u}_i(k)$ and $\Delta\tilde{u}_i(k)$.\\
  Update $\tilde{y}_i(k+1)$ and $\Delta\tilde{y}_i(k+1)$.\\
  \tcp{CPL}
    Update $\hat{\phi_i}(k)$, $u_i(k)$ and $\Delta u_i(k)$.\\
  Update ${\Delta \hat{ \chi}_i(k)}$.\\
  Apply $u_i(k)$ and $i=i+1$.\\
  }
 $k=k+1$.
  }
 \end{algorithm}

\
\section{ Numerical Simulation }\label{section5}
In this section, we show the effectiveness of the aforementioned theoretical result using a simulated example.

\begin{figure} 
    \centering
    \includegraphics[width=6.5cm]{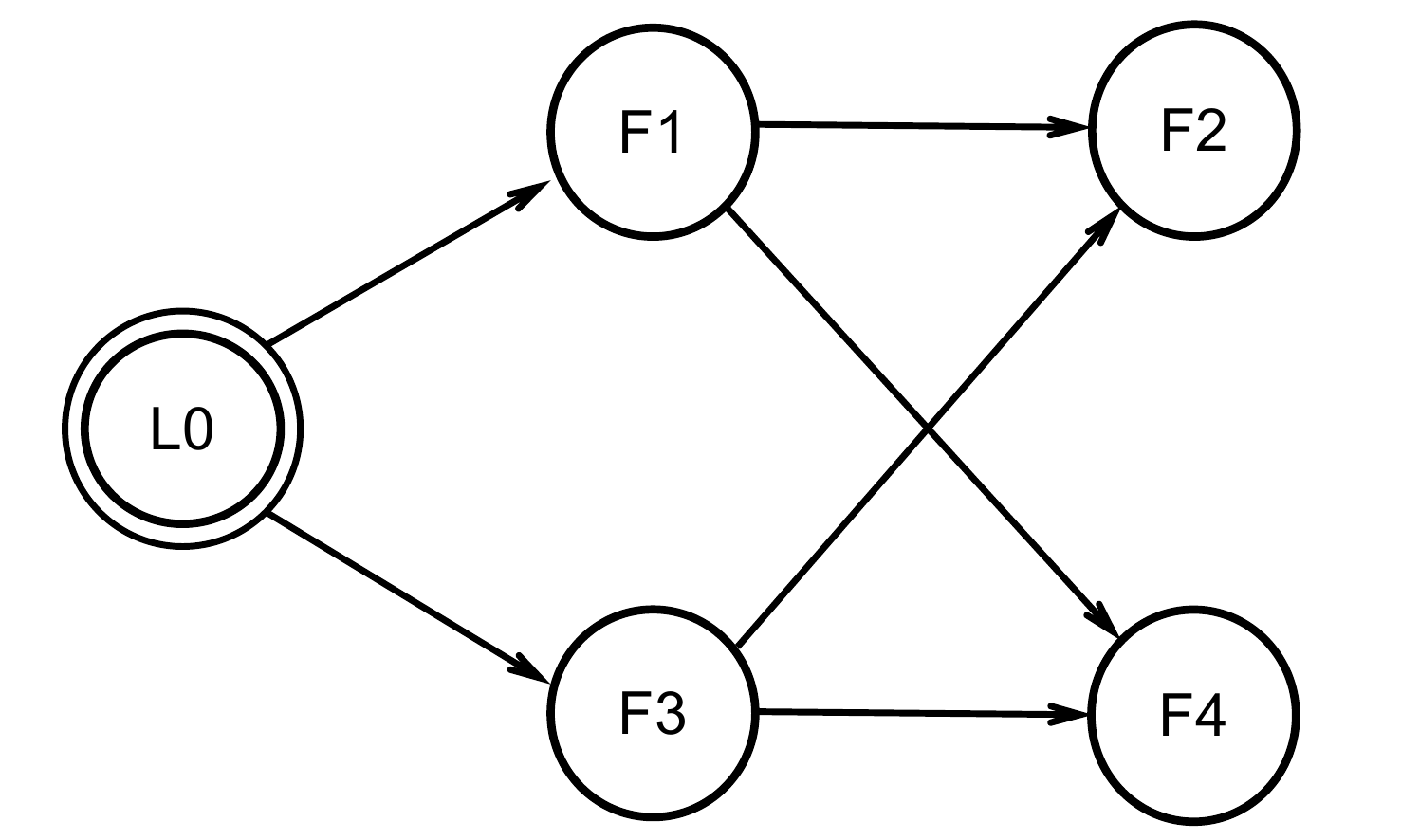}
    \caption{Communication topology.}\label{L}
\end{figure}

Considering a nonlinear MAS with one leader and four followers, the communication topology satisfying Assumption \ref{as_1} is shown in Fig. \ref{L}, in which the leader is denoted as node $0$ and four followers are denoted as node $1,\cdots, 4$. As the illustration shown in Fig. \ref{L}, the leader $0$ can only communicate with followers $1$ and $3$, transmitting own state information to them. The followers $2$ and $4$ can not receive the leader information, but it could receive the information from agents $1$ and $3$ to finish the tracking task due to that the communication graph is strongly connected. According to Communication topology in Fig. \ref{L}, the Laplacian matrix of the graph is obtained as
\begin{equation*}
   L= \left[
  \begin{array}{cccc}
 1 &0 &0 &-1\\
-1 &2 &-1 &0\\
0 &-1 &1 &0\\
-1 &0 &-1 &2\\
  \end{array}
  \right],
\end{equation*}
and $C={\rm diag}(1,0, 1 ,0)$.

Followers only directly or indirectly receive the state information of leader to complete leader-following consensus control, that is, only the leader state output is concerned in the process of control. Hence, autonomous system model of leader is represented in the form of state trajectory as follows:
\begin{equation*}
   y_0(k)= \sin{\frac{\pi k}{100}}+0.5 \cos{\frac{\pi k}{40}}.
\end{equation*}

The nonlinear dynamics of four follower agents 
satisfying Assumption \ref{as_2}-\ref{as_4} are shown as follows
\begin{equation*}
    \begin{aligned}
    \textbf{Agent 1: } y_1(k+1)&=\frac{|y_1(k)| u_1(k)}{1+{y_1(k)}^2}+0.2u_1(k).\\
    \textbf{Agent 2: } y_2(k+1)&=\frac{|y_2(k)| u_2(k)}{1+{y_2(k)}^3}+0.9 u_2(k).\\
    \textbf{Agent 3: } y_3(k+1)&=\frac{|y_3(k)| u_3(k)}{1+{y_3(k)}^2}+0.5u_3(k).\\
    \textbf{Agent 4: } y_4(k+1)&=\frac{|y_4(k)| u_4(k)}{1+{y_4(k)}^5}+0.8u_4(k).
    \end{aligned}
\end{equation*}

\subsection{Performance of TL against DoS Attacks}

The virtual followers on the TL is designed as
\begin{equation*}
    \begin{aligned}
    \textbf{Virtual Agent 1: } \tilde{y}_1(k+1)&=\frac{|\tilde{y}_1(k)| \tilde{u}_1(k)}{1+{\tilde{y}_1(k)}^2}+\tilde{u}_1(k).\\
    \textbf{Virtual Agent 2: } \tilde{y}_2(k+1)&=\frac{|\tilde{y}_2(k)| \tilde{u}_2(k)}{2+{\tilde{y}_2(k)}^3}+0.5 \tilde{u}_2(k).\\
    \textbf{Virtual Agent 3: } \tilde{y}_3(k+1)&=\frac{|\tilde{y}_3(k)| \tilde{u}_3(k)}{1+{\tilde{y}_3(k)}^3}+0.8\tilde{u}_3(k).\\
    \textbf{Virtual Agent 4: } \tilde{y}_4(k+1)&=\frac{|\tilde{y}_4(k)| \tilde{u}_4(k)}{1+{\tilde{y}_4(k)}^4}+\tilde{u}_4(k).
    \end{aligned}.
\end{equation*}

The nonlinear dynamics of virtual followers on the TL above are satisfying Assumption \ref{as_7}, and the boundary of $\Phi_i(k)$ is obtained as $bt=1.5$ according to dynamics model above. The initial state and initial input of virtual followers are selected as $\tilde{Y}(0)={[0.1 \,0.2 \,0.2 \,0.3]}^{\mathrm{T}}$ and  $\tilde{U}(0)={[0\, 0\, 0 \,0]}^{\mathrm{T}}$, respectively. The parameters of the DMFAC algorithm are chosen as $\eta_t=1$, $\mu_t=1$, $\gamma_t=0.6$ and $\lambda_t=1$, satisfying the condition of Theorem \ref{Th_1}. 
In this simulation, the DoS attacks occurs randomly in the whole control process, and the corresponding DoS attacks parameters are as follows: $M=10$ and $\beta=0.2$.

Fig. \ref{F_1} illustrates the virtual followers tracking trajectory with the DMFAC algorithm against DoS attack on the TL.
In order to observe the tracking effect after the DoS attacks, the DoS attacks is mainly concentrated in the first part of the control process as shown in Fig. \ref{F_1} (red areas indicate the occurrence of DoS attacks).
 According to the convergence rate of the tracking error at each time in the simulation results, we can get the convergence rates $\alpha_1=0.9$ and $\alpha_2=1.05$ in two cases, which proves that the choice of DoS attack parameters satisfies the condition of Theorem 1. 

As can be seen from the local enlarged image in Fig. \ref{F_1}, it is a process of estimating $\hat{\Phi}_i$ of DMFAC at the beginning, which is accompanied by large tracking fluctuations.
Although the virtual followers on the TL will maintain the original control input and deviate from the leader when the DoS attacks occur, it can speed up to follow the leader when the attacks are over.
Tracking error on the TL, which is UUB convergence with $B_t=0.32$, obviously, is illustrated in Fig. \ref{F_2}
 (yellow area is the boundary range).

\subsection{Performance of CPL with Double-layer DMFAC Algorithm}
Then we focus on the performance of CPL against both DoS attacks and unbounded AAs under double-layer DMFAC framework.


The AAs is chosen as: $\chi_1(k)=0.01k$, $\chi_2(k)=0.02k$, $\chi_1(k)=-0.01k$ and $\chi_1(k)=-0.02k$, variation of which are bounded by $\overline{d}=0.03$. According to the nonlinear dynamics of actual followers, we can get
$bc=1.5$. The initial state and and initial input of followers are $Y(0)={[0\, 0 \,0 \,0]}^{\mathrm{T}}$ and  $U(0)={[0\, 0\, 0\, 0]}^{\mathrm{T}}$, respectively. In order to meet the conditions of Theorem \ref{Th_2}, DMFAC parameters on the CPL are selected as $\eta_c=1$, $\mu_t=c$, $\gamma_c=0.8$ and $\lambda_t=c$. 

Combined with DMFAC on the TL, the effect of the double-layer DMFAC algorithm is shown in the Fig. \ref{F_3} and Fig. \ref{F_4} below.
The trajectories of all agents on the CPL are depicted in Fig. \ref{F_3} while the tracking error of followers is shown in Fig. \ref{F_4}.
Same as track on the TL above, there is a fluctuation adjustment at the beginning of the control process. In this process, DMFAC algorithm iteratively calculates the DMFAC parameters $\hat{\phi}_i$ and AAs attack compensation value $\Delta \hat{\chi}_i$. In addition, as can be seen from the partial enlarged view, when DoS attacks occur, the tracking effect will be disturbed, but after the attacks, it can quickly achieve the desired tracking effect. In addition, the tracking error will fluctuate slightly in the middle of the control process, which is related to the leader's state Variation. At this moment, the leader's state Variation is at its maximum, and the PDD parameter estimation $\hat{\phi}_i$ in DMFAC is also changing rapidly, so there is a small fluctuation.
Besides, the range of a tracking error is also shown in the yellow area of the Fig. \ref{F_4}
with upper bound $B=0.38$.

\section{Conclusion}
The leader-following consensus control for unknown nonlinear MASs against requency-constrained DoS attacks and unbounded AAs has been solved in this paper. A double-layer DMFAC framework based on the TL is proposed, which not only solves the problem of unknown nonlinearity with FDL-MFAC, but also resists both DoS attacks and AAs in different defense strategies. Strict proof is presented to guarantee that the tracking error is UUB convergence, which is verified by simulation above. Future works will consider the privacy of TL against other attacks, or the ability to solve other attacks except DoS attacks and AAs by building a multi-layer control framework.


\bibliography{ref}

 \end{document}